\documentstyle[epsfig]{article}
\begin{document}

\title{Are radio pulsars strange stars ?}
\author{R.C.Kapoor\thanks{email : rck@iiap.ernet.in}\\
 Indian Institute Of Astrophysics, \\Koramangala,
 Bangalore, 560 034, India  \and
       C. S. Shukre\thanks{email : shukre@rri.res.in}\\
 Raman Research Institute, \\ Sadashivanagar,
 Bangalore, 560 080, India}

\date{\today}
\maketitle

\begin{abstract}
A remarkably precise observational relation for pulse core component
widths of radio pulsars is used to derive stringent limits on pulsar
radii, strongly indicating that pulsars are strange stars rather than
neutron stars. This is achieved by inclusion of general relativistic
effects due to the pulsar mass on the size of the emission region
needed to explain the observed pulse widths, which constrain the
pulsar masses to be $\le \; 2.5 \; M_\odot$ and radii $\le \; 10.5 \;km$.
\end{abstract}

\section{Introduction}

Radio pulsars are believed to be the most common manifestations of
neutron stars, but it has not been possible so far to relate the
voluminous data on radio pulses and their varied structure to the
properties of neutron stars except through the arrival times of pulses.
Here we make such a connection between pulse core component widths
derived from very good quality radio data and the mass-radius (M-R)
relation of neutron stars. This becomes possible only due to the
inclusion of general relativistic effects of the stellar mass on pulsar
beam shapes, which makes the stellar mass and radius relevant parameters
in determining the pulse widths. We show that core component widths
provide tight constraints on equations of state (EOS) of neutron stars.
We compare our results with other similar attempts based, e.g., on the
X-ray data. From our constraints it emerges that no neutron star EOS seem
to be adequate, leading to the conclusion that pulsars are strange stars,
i.e., ones composed of quarks of flavors u, d, and s (Alcock et al. 
1986); and we examine it in light of similar recent suggestions.

\section{Core component widths}

A classification of radio pulse components into `core' and `conal'
emissions has emerged which is based on various characteristics such as
morphology, polarization, spectral index etc. of the pulses (Rankin, 1983).
Radio pulsars often show a three peaked pulse profile, the central
component of which is identified as the core emission, as opposed to the
outrider conal pair (Rankin, 1990). By analysing the core components of
many pulsars, especially the `interpulsars' which emit two pulses half a
period apart in one pulse period, Rankin (1990) found a remarkable
relation between the pulse width W and the pulsar period P (in seconds)
for pulsars whose magnetic dipole and rotation axes are orthogonal, viz.
\begin{equation}
W = \frac {2.^o45} {\sqrt P} \; \;  for \; \; \alpha = \pi/2
\label{E1}
\end{equation}
Here $\alpha$ is the angle between magnetic and rotation axes. This
relation (henceforth the Rankin relation) provides a fit to data
within $\simeq 0.2$ \% and the observations themselves have errors on
the average of $\simeq 4$ \%. Thus Eq. 1 is a rare example of an
extraordinarily good fit. In addition, the Rankin relation has also been
used (Rankin 1990) to predict $\alpha$ values for some other pulsars
which are not interpulsars. These predicted values agree very well with
determinations of $\alpha$ based on data about other components in the
same pulsars (Rankin 1993). Thus its remarkable fit to the core component
data is supported in addition by data on other pulsars.  The Rankin
relation in our view is one of the most reliable observational relations
derived from the radio pulsar data.

The import of the currently accepted `polar cap model' of pulsar radio
emission is that the radiation originates from the magnetic polar
regions. The polar cap is defined on the stellar surface by the feet
of the dipolar magnetic field lines which penetrate the `light cylinder',
i.e. a cylinder of radius $c P / 2 \pi$ with rotation axis as its
axis. $c$ is the speed of light. Pulsar emission occurs in this `open
field line' flux tube at an altitude $r$ measured radially from the
center of the star. We refer to the surface of emission as the \emph{
emission cap} which coincides with the polar cap when $r=R_*$ the
stellar radius.

As shown in Fig. 1 the line of sight cuts the polar cap along the line LS.
This will lead to a pulse of width $W$. If LS passes through the centre,
then $W=\, 2\, \rho$, the longitudinal diameter of the polar cap. For
interpulsars LS passes very close to the centre and hence $W \simeq 2 \,
\rho$.
For a value of  $\alpha \ne 90^\circ$, $2\,\rho$ can not be recovered
from $W$ alone. One also needs to know the displacement of LS from the
centre, usually called the impact angle $\beta$. If polarization data
is available in addition to $W$, then both $\beta$ and $2\,\rho$ can be
retrieved from observations. The core component width data used by
Rankin (1990) pertains only to inter-pulsars. Therefore, in essence the
width W in Eq. \ref{E1} is the longitudinal diameter $2 \rho$ of the
emission cap and is thus independent of $\alpha$ (Kapoor and Shukre,
1998, henceforth KS). From the dipole geometry (Goldreich and Julian,
1969, henceforth GJ) one finds
\begin{equation}
2 \rho = \frac {2.^o49} {\sqrt P}  \sqrt{ \frac {r} {10\,km},
}\label{E2}
\end{equation}

\begin{figure}
\epsfig{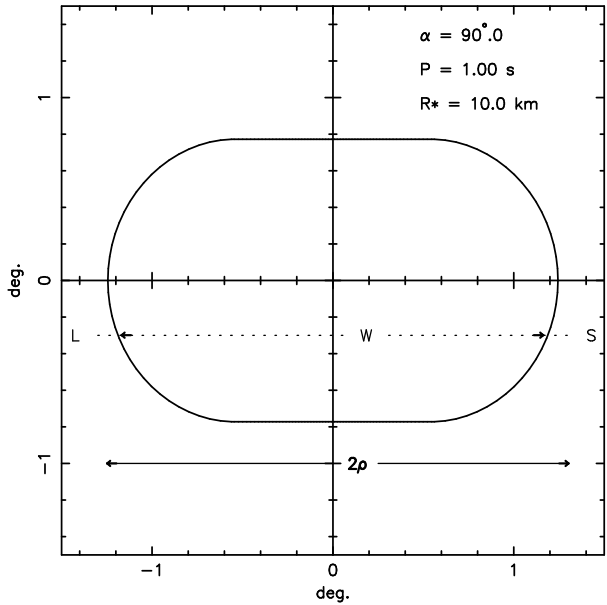}
\caption{Polar plot of the polar cap of a typical pulsar. The $x$ and $y$
co-ordinates are the magnetic longitude and lattitude respectively. The
centre of the figure represents the radial direction passing through the
dipole magnetic axis. Dotted line LS shows the locus of the line of sight
as the pulsar rotates. See text for further details.}
\label{f1}
\end{figure}

On the assumption that the full emission cap participates in the
core emission, agreement between Eqs. \ref{E1} and \ref{E2} immediately
allows the conclusion that $r = 10\, km$. This remarkable agreement
has provided compelling evidence favouring the origin of the core
emission from the stellar surface as well as the dipolar configuration
of the stellar magnetic field (Rankin, 1993). Note that a value of the
stellar radius $R_*$ really has not entered the considerations so far.
However, $10 \, km$ is considered to be the cannonical value of $R_*$,
and it is on this basis that $r$ is identified with $R_*$.

\section{General relativistic widths and constraints on pulsar mass
 and radius}

In the analysis of radio pulse structure, if the role of the radius $R_*$
has been insignificant, then it is even more so for the stellar mass $M$.
Inclusion of effects due to the spacetime curvature caused by pulsar's
mass changes this as follows. The stellar gravitational field affects the
dipole field geometry and also causes bending of the rays of the emitted
pulsar radiation. The former tries to shrink the emission cap while the
latter has the opposite effect of widening it. A detailed study of these
effects has been done and described in KS. In summary, we give below an
analytic but approximate version of how Eq. \ref{E2} is modified, i.e.,
\begin{equation}
2 \rho = \frac {2.^o49} {\sqrt P} \; \sqrt{ \frac {r} {10\,km} } \;
f_{sqz} \;f_{bnd},
\label{E3}
\end{equation}
where the factors $f_{sqz}$ and $f_{bnd}$ are respectively due to
squeezing of the dipole magnetic field and bending of light by the
stellar gravitation and are given by
\begin{equation}
f_{sqz} = (1+ \frac {3m} {2r} )^{ -\frac {1} {2}}, \hspace{0.4cm}
f_{bnd} = \frac {1}{3} ( 2 + \frac {1} {\sqrt{1-\frac {2m}{r} }
} ),
\label{E4}
\end{equation}
where $m= \frac {G\,M}{c^2}$, i.e., $2\,m$ is the Schwarzshild radius.
Eq. \ref{E2} is recovered in the limit $m=0$.

In Eq. \ref{E3} the effects due to special relativistic aberration are
not included. Since stellar gravitational effects are significant for
$r\le \;20\,m$ (KS) we consider only such emission altitudes here. Even
for the $1.5 \,ms$ pulsar PSR 1929+214, therefore, aberration does not
play a role in considerations here. In what follows, however, the
calculations include all the effects completely, as in KS. For $M=1.4\,
M_\odot$ and $R_*=\, 10\,km$, the net effect on the emission cap on the
surface is a shrinking by $\sim 4$ \% compared to the value in Eq.
\ref{E2}. Although small, this difference allows us to relate $M$ and
$R_*$, and as we shall see provides tight constraints on the pulsar EOS.

Fig. 2 shows the variation of $2\rho$ with $r$ for various values of $M$
as labelled. The points where the Rankin line intersects the curve for a
particular mass $M$ gives for that $M$ the altitude(s) where the core
emission must originate.

Generally there are two intersection points, $r_1$ and $r_2$, such that
$r_1 \le r_2$. In the limiting case $M=M_0$ the two points coalesce. For
higher values of $M$ there is no intersection. The mass $M_0$ is $2.48 \,
M_\odot$, which we take as $2.5 \, M_\odot$. Thus we can conclude that
core emission does not occur if $M \, > \, M_0$. Probably, this is an
indication that \emph{all} radio pulsars have masses $< \, M_0$ because
the incidence of core emission among radio pulsars is $\sim 70$ \%
(Rankin, 1990). Thus
\begin{equation}
M \le M_0 \simeq 2.5 M_\odot. \label{E5}
\end{equation}
This constraint, though of interest, is not useful since observationally
all masses seem to be well below it.

The second constraint involves $R_*$. The lowest altitude at which any
emission can occur is $R_*$. Therefore for values of $M$ below $M_0$,
\begin{equation}
R_* \le r_1 \; and/or \; r_2. \label{E6}
\end{equation}
Since $r_1$ and $r_2$ depend on $M$ we get a constraint on the pulsar
mass-radius relation from the inequalities \ref{E6}.

\begin{figure}
\epsfig{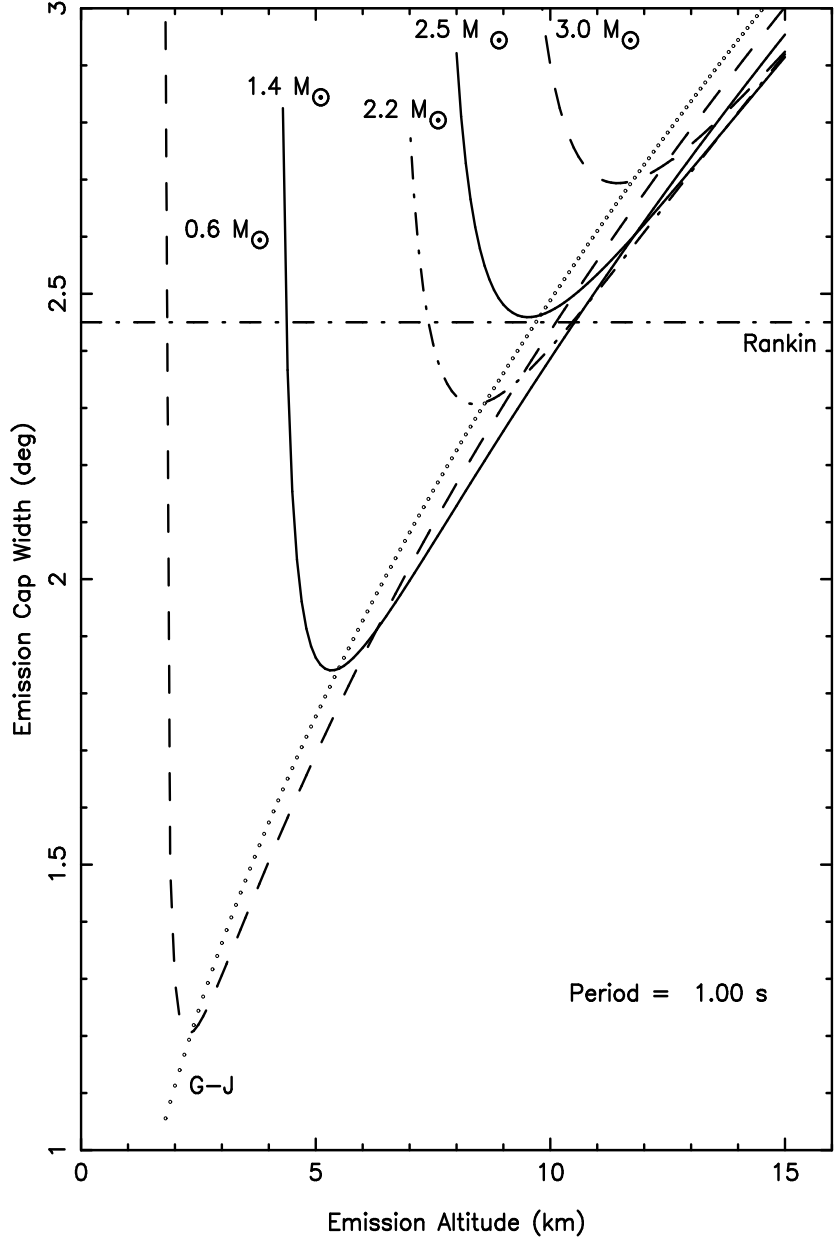}
\caption{Width $2 \rho$ of the emission cap after inclusion of all
special and general relativistic effects \emph{vs.} the emission altitude
$r$ for various stellar masses as shown. Horizontal line is the Rankin
relation of Eq. \ref{E1} and the dotted line labelled GJ is given by Eq.
\ref{E2}. For a different pulsar period, widths scale as ${\sqrt P}$ as
in Eq. \ref{E3}.}
\label{f2}
\end{figure}

\begin{figure}
\epsfig{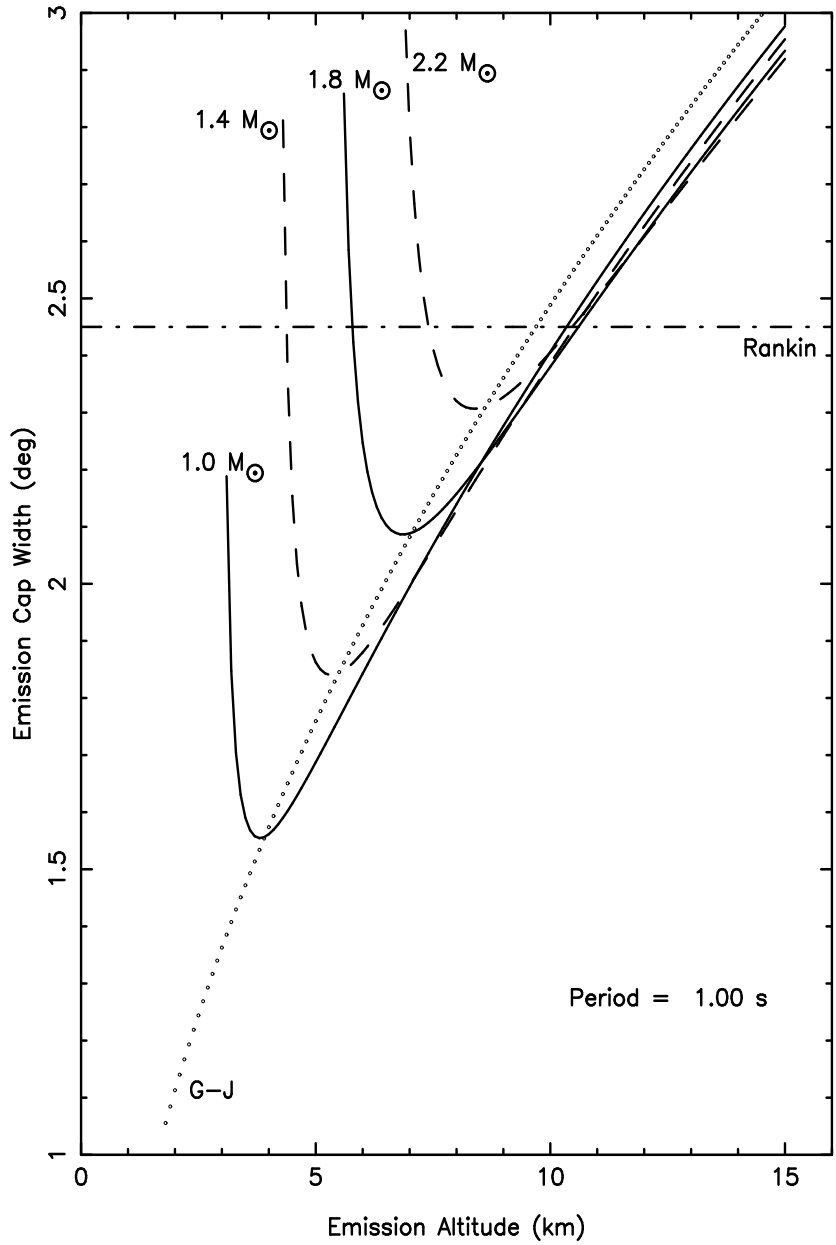}
\caption{Same as Fig. \ref{f2} but for $1.0\,M_\odot<M<2.2\,M_\odot$.}
\label{f3}
\end{figure}

For all masses, values of $r_1$ are almost same as $2\,m$ and if taken
seriously would imply that pulsars are black holes. We therefore consider
only $r_2$. Values of $r_2$ range from 10.2 to 10.6 $km$ for masses
between 0.6 to 2.5 $M_\odot$. For masses between 1 $M_\odot$ and 2.2
$M_\odot$, values of $r_2$ remarkably enough are not very sensitive to
$M$ and are all close to 10.5 $km$ as seen in Fig. \ref{f3}. Again, pulsar
masses are observationally seen to be well covered by the range 1.0 - 2.2
$M_\odot$ and so we can take $10.5\;km$ as the upper limit for all $M$.
Lower values of $r_2$ occur for lower values of M and their inclusion will
only tighten the constraint further since for all EOS a decrease in mass
implies an increase in radius. The Rankin relation thus leads us to the
second constraint
\begin{equation}
R_* \, \le  \; 10.5 \, km, \label{E7}
\end{equation}
which is applicable to radio pulsars which show core emission, and, as
remarked earlier, to most probably all pulsars.

\begin{table}[h]
\caption {EOS for which M - R plots are available. For meaning of
$M_{min}$ and $M_{max}$ see text.}
\label{tabl1}
\begin{tabular}{rcllrrcllr}
   &        &           &             &      &
   &        &           &             &      \\ \hline
SN & EOS    & $M_{min}$ & $M_{max}$   & Plot &
SN & EOS    & $M_{min}$ & $M_{max}$   & Plot \\
   & Name   &($M_{\odot}$)& ($M_{\odot}$) & Ref.$^1$ & 
   & Name   &($M_{\odot}$)& ($M_{\odot}$) & Ref.$^1$ \\ \hline
 1 & A      & 0.35 & 1.65 & BBF,PC    &
12 & Hyp    & - -  & - -  & LRD \\
 2 & B      & 0.35 & 1.40 & BBF       &
13 & BPAL12 & 1.35 & 1.45 & LRD \\
 3 & M      & - -  & - -  & PC        &
14 & BBB1   & 1.65 & 1.75 & LRD \\
 4 & L      & - -  & - -  & BBF,PC,BLC&
15 & BBB2   & 1.70 & 1.90 & LRD \\
 5 & WFFAU  & 0.45 & 2.15 & BLC       &
16 & EOS1   & 1.50 & 1.55 & BLC \\
 6 & WFFUU  & 2.10 & 2.20 & PC,LRD    &
17 & EOS2   & 1.70 & 1.75 & BLC \\
 7 & WFFUT  & 1.65 & 1.85 & BBF       &
18 & RH     & 0.15 & 0.90 & HWW \\
 8 & FPS    & 1.60 & 1.80 & PC,BLC    &
19 & RHF    & 0.15 & 0.95 & HWW \\
 9 & HV     & - -  & - -  & W         &
20 & APR1   & - -  & - -  & BBF \\
10 & HFV    & - -  & - -  & W         &
21 & APR2   & 2.15 & 2.2  & BBF \\
11 & G$^{\pi}_{300}$&- -&- - & W      &
22 & $K^{-}$& - -  & -  - & LRD \\ \hline
\end{tabular}
{$^{1}$References are same as in the reference section at the end:
BBF-Benhar et al. 1999; PC-Psaltis \& Chakrabarty 1999; BLC-Balberg et
al. 1999; LRD-Li et al. 1999b; W-Weber 1999; HWW-Huber et al. 1998}
\end{table}

\section{Constraints and neutron star EOS}

We have searched earlier works for neutron star M - R relations. For about
40 EOS M - R plots were available. Very conservatively dropping some
among them which are now replaced by modern versions, we have selected
the 22 listed in Table \ref{tabl1}. For the additional six in Table
\ref{tabl2} only the maximum masses ($M_{max}$) allowed by the EOS and
the associated radii are available (Salgado et al. 1994).

For all EOS in Table \ref{tabl2}, radii are larger than $10.5 \; km$ for
$M \, = \, M_{max}$ and thus also for lower values of $M$. Therefore we
consider now the 22 remaining EOS in Table \ref{tabl1}. Since high
precision is not called for, or, is available, we have read off from the
published plots the mass range for which $R_* \,< \, 10.5 \, km$. These
values are listed in Table \ref{tabl1} as $M_{min}$ - the mass for which
R $= \, 10.5 \, km$ and $M_{max}$ - the maximum mass allowed by the EOS.
Where the EOS does not permit $R_*$ $< \, 10.5 \, km$ for any mass, only
dashes appear for $M_{min}$ and $M_{max}$. There are 8 such EOSs and they
are not favored by inequality \ref{E7}.

Because of the accurately determined masses for the Hulse-Taylor binary
system (i.e., 1.44 and 1.39 $M_\odot$) (Thorsett and Chakrabarty, 1999),
for the remaining EOS we impose an additional condition that their mass
range allow the value 1.4 $M_\odot$. The inequality \ref{E7} selects out
the softer EOS. By imposing this condition based on observations we are
in effect demanding that the EOS should not be so soft as to have
$M_{max} \, < \, 1.4 \, M_\odot$ or so stiff that $M_{min} \, > \, 1.4 \,
M_\odot$. This further reduces the number of acceptable EOSs by 11. The
remaining three are : A, WFFAU and BPAL12.

The core width constraints in conjunction with the observational
information on pulsar masses have thus reduced the viable netron star
EOS number from 28 to 3.

The EOS APR1 is an updated version of the EOS A. APR2 is APR1 with
relativistic corrections included. Since both APR1 and APR2 do not
survive the constraints we can drop also the EOS A from the short list.
In addition, based on general restrictions following from the glitch data,
Balberg et al. (1999) have disqualified the EOS A and WFFAU. We are thus
left with the choice of BPAL12 or some variant of it as the only viable
modern EOS. 

\begin{table}
\caption {EOS for which only M$_{max}$ and its radius R are available.}
\label{tabl2}
\begin{tabular}{rcccrccc}
   &&&&&&&\\ \hline
SN & EOS$^1$ & M$_{max}$$^1$ & R(M$_{max}$)$^1$&SN & EOS$^1$ & M$_{max}$$^1$&
R(M$_{max}$)$^1$ \\
& Name  & $M_{\odot}$ & $km$ & &Name  & $M_{\odot}$ & $km$ \\ \hline
 1 & HKP    &  2.83   & 13.68 & 4 & Glend3 &  1.96   & 11.30 \\
 2 & Glend1 &  1.80   & 11.15 & 5 & DiazII &  1.93   & 10.93 \\
 3 & Glend2 &  1.78   & 11.29 & 6 & WGW    &  1.97   & 10.97 \\ \hline
\end{tabular}
\\
{$^{1}$Names and values are from Salgado et al. (1994).}
\end{table}
We have considered only the non-rotating neutron star models because most
pulsars are slow rotators. But inclusion of rotation (or magnetic field)
will not change the situation because, in that case, for a given mass one
expects larger radii on general physical grounds.

It should be noted that similar attempts using the pulsar timing data
(glitches) and X-ray source data (quasi-periodic oscillations) do not
provide such stringent constraints and are also not so selective of the
EOS (Psaltis and Chakrabarty, 1999, van Kerkwijk et al., 1995). Also, our
constraints are not dependent on uncertainties in theoretical models,
i.e., of accretion disks, and rely on very simple and fundamental
assumptions.

Our constraints make crucial use of the Rankin relation and the
assumption that the core emission emanates from the full polar cap.
\emph{It will be of great interest to re-evaluate both of these
independently}. The database presently available is presumably more
voluminous than in 1990 because the number of known pulsars has more
than doubled since then and it can be used to further fortify the Rankin
relation. On the other hand it would be worthwhile also to check the
assumption of the participation of the full cap by some independent means.
Non-dipolar magnetic field components have been invoked in the past in
various contexts (Arons, 2000, Gil and Mitra 2000). Our analysis
crucially hinges on the Rankin relation, which in turn makes crucial
use of the dipole nature of the field. The existence of non-dipolar
components has been studied by Arons (1993) and he has concluded
against their presence. We take the view that the remarkable agreement of
the Rankin relation actually provides evidence for the dipolar nature of
the field and strongly indicates the absence of non-dipolar components
and also of propagation effects affecting the core emission.

\section{Are radio pulsars strange?}

In so far as our constraints hold, can we then conclude that BPAL12 is
{\it the} neutron star EOS ? Actually BPAL12 is used as an extreme case
for illustrative purpose and can hardly be called a realistic netron star
EOS (Bombaci 2000). In fact our present knowledge of the neutron star EOS
is very far from final. Present theoretical uncertainties in these EOS
relate to the very high density regime ( $\rho \,  \sim \, 10^{15} \,
gm \, cm^{-3}$ ) and are small in terms of pressure. For our constraint,
however, these small changes in pressure are significant and can lead to
very different radii $R_*$ (See Figs. 2 and 3 in Benhar et al. 1999). The
best we can do is to glean from the trend which is visible in the EOS that
include the microphysics in the best possible way, i.e., those based on
relativistic quantum field theory (Salgado et al., 1994, Prakash et al.,
1997), rather than those in which nucleon interactions are described
using potentials (as in the BPAL series). These are the EOS in Table
\ref{tabl2} and none among these theoretically most advanced EOS are
favored by our constraints. (This is also true of similar EOS described
in Prakash et al. 1997). Extrapolating on this trend it would seem that
no neutron star EOS can satisfy the inequality \ref{E7}. This in turn
implies that pulsars are not neutron stars\footnote{ Recently, based on
general and well-accepted principles it has been shown (Glendenning, 2000)
that it is possible to have small radii for neutron stars, but none of
the known EOS show this. Interestingly, for a radius $< \, 10.5 \, km$
the maximum mass turns out to be $2.5 \, M_\odot$, in close agreement
with the inequality \ref{E5}. } and leaves us with the only alternative
conceivable at present, that pulsars are strange quark stars. We discuss
this next.

Some stars considered so far to be neutron stars have been proposed to be
actually strange stars on two counts. The proposals for Her X-1 (Dey et
al. 1998), 4U 1820-30 (Bombaci, 1997),  SAX J1808.4-3658 (Li et al.,
1999a), 4U 1728-34 (Li et al., 1999b) are based on the compactness of
stars being more than a neutron star can accomodate. From an entirely
different viewpoint PSR 0943+10 has been proposed to be a bare strange
star (Xu et al., 1999). This last proposal implies that all pulsars
showing the phenomenon of drifting subpulses may be bare strange stars.

Pulsars being strange stars fits well with our constraints. Whether
pulsars are bare strange stars, strange stars with normal crusts or the
newly proposed third family of ultra-compact stars (Glendenning and
Kettner, 2000) is difficult to decide at present. For the relatively
better-studied strange stars, the new EOS for strange stars give radii
$\simeq\, 7 \, km$ as opposed to $\simeq \, 8 \, km$ given by earlier EOS
based on the MIT bag model (Dey et al., 1999). Xu et al.(1999) propose
that pulsars showing the phenomenon of drifting sub-pulses are bare
strange stars. Our constraints apply to pulsars showing core emission.
However, the core emission and drifting of subpulses which is a property
of the conal emission (Rankin, 1993; see also  Xu et al., 1999) are not
mutually exclusive. Therefore the proposal that pulsars are bare strange
stars can be extended to all pulsars. Many issues, such as differences
between bare strange stars and those with normal crusts etc. remain to be
answered, although some answers have been proposed. We do not repeat here
this discussion  (Xu et al., 1999, Madsen, 1999) except to state that our
core width constraints are one more independent indication that pulsars
are strange stars.

The source SAX J1808.4-3658 has been proposed to be a strange star on the
basis of its compactness. In the analysis of Psaltis and Chakrabarty
(1999) it was demonstrated that the presence of multipole components
relaxes the amount of compactness required, such that the star could be a
neutron star. In our analysis also, existence of multipoles (however
{\it ad hoc}) would dilute our conclusion of pulsars being strange stars.
It thus seems that existence of multipoles or the strange star nature of
hitherto considered neutron stars are two mutually exclusive choices. At
present it is very difficult to choose between them. More work on strange
stars may in future elucidate this, but introduction of multipoles brings
in so many parameters that how their existence could be proved from
observations is unclear. The multipoles would also rob the Rankin
relation of its beauty and turn its remarkable observational agreement
into a mystery.

\section{Summary}

In summary, the empirical formula of Rankin (1990) describing the
opening angle of the pulsar beam emitting the core emission when compared
to theoretically calculated value leads to a constraint that pulsar
masses should be $\le \; 2.5 \; M_\odot$ and radii $\le \; 10.5 \; km$.
This comes about due to the inclusion of general relativistic effects of
the mass of the star on the pulsar beam size. For observationally
reasonable pulsar masses a comparison with mass-radius relations of
neutron star EOS shows that most of the EOS are ruled out, implying that
pulsars are strange stars and not neutron stars, unless our understanding
of the neutron star EOS is revised.

{\it Acknowledgements}
One of us (C.S.S.) would like to thank Rajeev Gavai and Sanjay Reddy for
rekindling his interest in the subject matter of this paper whose
preliminary version was reported in Kapoor and Shukre (1997). We thank
I. Bombaci and Joanna Rankin for valuable comments.

\end{document}